\documentclass{jkas}

\def\year{2023} 
\def\volume{56} 
\def\issue{2} 
\def\beginpage{1} 
\def\received{November 24, 2023} 
\def\accepted{December 21, 2023} 
\def\published{December 00, 2023} 
\date{Received \received; Accepted \accepted; Published \published}

\title{%
Number of Scatterings in Random Walks
}

\author[$^{1,2\star}$]{Kwang-Il Seon}{0000-0001-9561-8134}
\author[$^{1,2}$]{Hyung-Joe Kwon}{}
\author[$^{1,2}$]{Hee-Gyeong Kim}{}
\author[$^{1,2}$]{Hyeon Jeong Youn}{}

\affil[$^1$]{Korea Astronomy \& Space Science Institute, 776 Daedeokdae-ro, Yuseong-gu, Daejeon 34055, Republic of Korea}
\affil[$^2$]{Astronomy and Space Science Major, University of Science and Technology, 217, Gajeong-ro, Yuseong-gu, Daejeon 34113, Republic of Korea}

\def\corrauthor{%
K.-I. Seon, \email{kiseon@kasi.re.kr}
}

\def\runningauthor{%
Seon et al.
}

\def\runningtitle{%
Number of Scatterings in Random Walks
}

\def\keywords{%
radiative transfer -- scattering -- methods: analytical
}

\def\abstracttext{%
This paper investigates the number of scatterings a photon undergoes in random walks before escaping from a medium. The number of scatterings in random walk processes is commonly approximated as $\tau+\tau^2$ in the literature, where $\tau$ is the optical thickness measured from the center of the medium. However, it is found that this formula is not accurate. In this study, analytical solutions in sphere and slab geometries are derived for both optically thin and optically thick limits, assuming isotropic scattering. These solutions are verified using Monte Carlo simulations. In the optically thick limit, the number of scatterings is found to be $0.5\tau^2$ and $1.5\tau^2$ in a sphere and slab, respectively. In the optically thin limit, the number of scatterings is $\approx\tau$ in a sphere and $\approx\tau(1-\gamma-\ln\tau+\tau)$ in a slab, where $\gamma\simeq 0.57722$ is the Euler-Mascheroni constant. Additionally, we present approximate formulas that reasonably reproduce the simulation results well in intermediate optical depths. These results are applicable to scattering processes that exhibit forward and backward symmetry, including both isotropic and Thomson scattering.
}

\begin{document}
\jkashead 


\section{Introduction\label{sec:intro}}
Scattering of radiation through a medium with a large optical depth occurs in  various astrophysical contexts. The most frequently encountered scattering mechanisms include Thomson scattering of photons by non-relativistic free electrons in ionized gaseous nebulae and resonance scattering of Ly$\alpha$ and Mg\,{\sc II} photons in the neutral interstellar and circumgalactic medium. In Thomson scattering, the scattering cross section is nearly independent of the incident photon's frequency. In earlier studies, the Ly$\alpha$ radiative transfer (RT) problem was also investigated under the assumption that the resonance cross section could be approximated as that at the line center, independent of the photon frequency \citep[e.g.,][]{Osterborck1962}, although, in reality, it is not.

The diffusion process of Thomson-scattered line radiation in a slab geometry, particularly in the context of polarization, was investigated in \citet{Chandrasekhar1960_book} and  \citet{Phillips1986}. The polarization of Thomson-scattered radiation in an oblate spheroidal medium was studied by \citet{Angel1969}. The wavelength dependence of polarization of a Thomson-scattered emission line was explored in \citet{Lee1999ApJ} and \citet{Kim2007MNRAS}.  A Monte Carlo study, conducted by \citet{Choe2023}, investigated the diffusion process of Thomson-scattered line radiation in both real space and frequency space.
Monte Carlo Ly$\alpha$ RT has also been extensively studied in many different astrophysical contexts \citep{Ahn2000JKAS,Chang2023,Gronke2014,Seon2020ApJS,Seon2022,Song2020,Verhamme2006,Yan2022,Zheng2002}.
Recently, a Monte Carlo RT simulation of the Mg\,{\sc II} $\lambda\lambda$2796, 2803 doublet line was also performed to investigate the doublet flux ratio \citep{Seon2023}.

\begin{figure*}[t]
\centering
\includegraphics[angle=0.0,width=125mm]{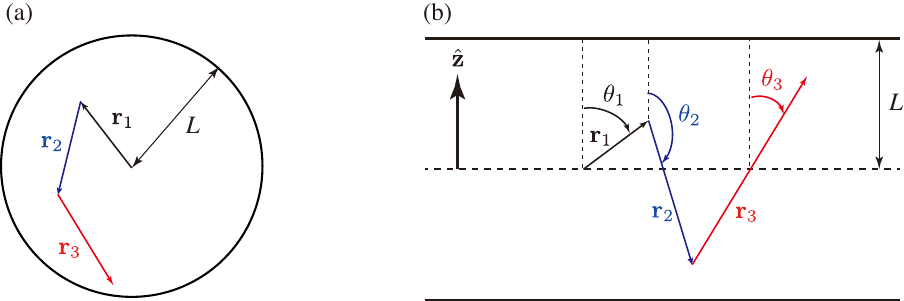}
\caption{Geometries adopted in this study: (a) sphere and (b) slab. The physical size ($L$) of the system is defined as the radius for the sphere and the height from the center for the slab geometry. The slab is infinitely extended along the $x$ and $y$ directions. In the slab geometry, the polar angle $\theta_i$ for the $i$th displacement vector is measured from the positive $z$ direction.\label{fig01}}
\end{figure*}

These diffusion processes lead to the broadening of emission lines, with their strength and width determined by the number of scatterings experienced by photons and the gas temperature. Therefore, one of the most critical factors in understanding line formation processes is the number of scatterings. \citet{Osterborck1962} applied the probability methods from the random walk process described in \citet{Chandrasekhar1943} to Ly$\alpha$ photons. They found that the mean number of scatterings required for a photon to diffuse out would be $\approx\tau_0^2$ in a medium with a very large optical depth, denoted as $\tau_0$. A similar argument was developed by \citet{Rybicki1986_book} for a random-walk process, demonstrating that the mean number of scatterings undergone by a photon before escaping an optically thick medium is $\approx\tau_0^2$.

In the Ly$\alpha$ RT process, the number of scatterings in the limit of large optical depth was analytically derived for a slab and a sphere by \citet{Harrington1973} and \citet{Seon2020ApJS}, respectively. \citet{Seon2020ApJS} validated the analytical solutions through Monte Carlo simulations, covering for a wide range of optical depths and gas temperatures in both slab and sphere geometries.
However, surprisingly, to the best of our knowledge, we found no literature validating the approximate formula of \citet{Rybicki1986_book} for the number of scatterings in the simple random walk process, where the scattering cross section is independent of the photon frequency.

This absence of validation for the approximate formula concerning the number of scatterings in random walks has motivated the present study. We soon recognized that the formula is not correct and, consequently, derived new analytical formulas for both a sphere and a slab. These results were further verified through Monte Carlo simulations. The analytical approximations are derived in Section \ref{sec:02}. Section \ref{sec:03} details the Monte Carlo simulation methods and results. The summary and discussion are found in Section \ref{sec:04}.

\section{Analytical Approximations\label{sec:02}}
The approach in this study is similar to that adopted in \citet{Rybicki1986_book}. However, we have derived the approximate equations accurately by carefully handling the probability distribution function for both sphere and slab geometries.

The scattering region is assumed to be a sphere or a slab, where photons are isotropically emitted from the center and scattering particles are uniformly distributed inside the region.
The optical thickness of a medium is given by $\tau_0=n\sigma L$, where $n$, $\sigma$, and $L$ represent the particle number density, cross section, and the radius of the sphere (or the height of the slab), respectively. The geometries and the definition of the system size $L$ are illustrated in Figure \ref{fig01}. The optical thickness can be expressed as $\tau_0=L/\ell$ in terms of the mean free path, $\ell=1/(n\sigma)$. In the following, we note that the probability of being scattered within an optical depth interval $d\tau$ at the optical depth $\tau$ is $P(\tau)d\tau=e^{-\tau}d\tau~~(0\le\tau \le \infty)$.

\subsection{Sphere}
In an optically thin sphere, photons will be scattered at most only once. Consequently, the mean number of scatterings eventually becomes equivalent to the probability that a photon undergoes scattering in the medium. The mean number of scatterings is then given by the following expression:

\begin{eqnarray}
N_{\rm scatt}\!\!\!&=&\!\!\!\int_0^{\tau_0} P(\tau)d\tau=1-e^{-\tau_0} \nonumber \\
\!\!\!&\!\!\!\approx&\tau_0 ~~~~(\tau_0\ll 1,~{\rm sphere}).\label{eq01}
\end{eqnarray}

For a medium of large optical thickness, photons will be scattered multiple times. The displacement of a photon between the $(i-1)$th and $i$th scatterings is denoted by $\mathbf{r}_i$, as illustrated in Figure \ref{fig01}. The net displacement of the photon after $N$ scatterings is then $\mathbf{R}=\mathbf{r}_1+\mathbf{r}_2+\mathbf{r}_3+\cdots+\mathbf{r}_N$. The average net displacement is obtained by squaring $\mathbf{R}$ and then averaging it over all photons:

\begin{eqnarray}
\label{eq:01}
\left<\mathbf{R}^2\right>=\sum_{i=1}^{N}\left<\mathbf{r}_i^2\right>+\sum_{\substack{i,j\\i\neq j}}^{N}\left<\mathbf{r}_i\cdot\mathbf{r}_j\right>,
\end{eqnarray}
where the angle brackets $\left<\right>$ denote the average over all photons. The second term in the equation involves averaging the cosines of angles between the directions before and after scatterings. In the summation, $i$ and $j$ do not necessarily denote sequentially occurring scattering events. In the optically thick limit, this will vanish for any scattering type with front-back symmetry, such as isotropic, Thomson, or Rayleigh scattering. Then, only the first term will contribute to the mean net displacement, which will consequently become $N_{\rm scatt}\left<\mathbf{r}^2\right>$, where $N_{\rm scatt}$ and $\left<\mathbf{r}^2\right>$ are the mean number of scatterings and the mean square displacement for a single scattering event, respectively. The mean square displacement of a single scattering event can be calculated as follows:
\begin{equation}
\left<\mathbf{r}^2\right>=\int_0^{\infty}\tau^2\ell^2 P(\tau)d\tau = 2 \ell^2
\end{equation}
The photon will escape after the symetem acquiring the net displacement of $N_{\rm scatt}\left<\mathbf{r}^2\right>=L^2$. Therefore, the mean number of scatterings undergone by photons is given by
\begin{equation}
N_{\rm scatt}\approx\frac{1}{2}\tau_0^2~~~~(\tau_0\gg 1,~{\rm sphere}).
\end{equation}

Combining the above equations for optically thin and thick cases, we obtain an approximation that might be applicable in both optically thin and optically thick media:
\begin{equation}
	N_{\rm scatt}^{\rm sphere}\approx\tau_0+\frac{1}{2}\tau_0^2.\label{eq-sphere}
\end{equation}

Our result differs from that of \citet{Rybicki1986_book} by a factor of 2 in the optically thick limit. This difference arises from their assumption that the path lengths between scatterings are always constant, rather than following the probability distribution function (PDF) $P(\tau)=e^{-\tau}$. Their result is not self-consistent because they assumed the PDF when deriving the number of scatterings in the optically thin limit but not in the optically thick limit.

\subsection{Slab}

In an optically thin slab, photons will escape the system when the $z$ component of the displacement vector expected at the first scattering event is beyond the half height of the slab, i.e., when $\left|\mathbf{r}_1\cdot\hat{\mathbf{z}}\right|>L$. Therefore, unlike in the spherical geometry case, the condition for scattering depends on the polar angle $\theta$ of the photon direction. The probability of being scattered when a photon is emitted into a direction of $\mu=\cos\theta$ is $p_{\rm scatt}(\mu)=1-e^{-\tau_0/\left|\mu\right|}$ because the optical thickness from the center to the slab boundary along that direction is $\tau_0/\left|\mu\right|$.
Therefore, the mean number of scatterings can be obtained by integrating the probability $p_{\rm scatt}(\mu)$ over $\mu$, which is uniformly distributed in the range of $-1\le\mu\le 1$. The resulting number of scatterings is then given as follows:
\begin{eqnarray}
N_{\rm scatt}\!\!\!&=&\!\!\!\frac{1}{2}\int_{-1}^1\left(1-e^{-\tau_0/\left|\mu\right|}\right)d\mu \nonumber\\
\!\!\!&=&\!\!\!1-e^{-\tau_0}+\tau_0\Gamma(0,\tau_0)~~~~(\tau_0\ll 1,~{\rm slab})\label{eq-for-thin-slab}
\end{eqnarray}
Here, $\Gamma(0,\tau_0)\approx -\gamma-\ln\tau_0+\tau_0+\mathcal{O}(\tau_0^2)$ is the upper incomplete gamma function, where $\gamma\simeq 0.57722$ is the Euler-Mascheroni constant. It is noteworthy that the number of scatterings for a slab is found to be higher than $N_{\rm scatt}\approx\tau_0$ obtained for a sphere. This is because the optical depth along a direction with $\theta>0$ is always higher than that at $\theta=0$.

In the optically thick limit, we need to consider the mean square of the $z$ component of the net displacement, which is given by:
\begin{eqnarray}
\left<\left(\mathbf{R}\cdot\hat{\mathbf{z}}\right)^2\right>\!\!\!&=&\!\!\!\sum_{i=1}^{N}\left<\left(\mathbf{r}_i\cdot\hat{\mathbf{z}}\right)^2\right>+\sum_{\substack{i,j\\i\neq j}}^{N}\left<\left(\mathbf{r}_i\cdot\hat{\mathbf{z}}\right)\left(\mathbf{r}_j\cdot\hat{\mathbf{z}}\right)\right> \nonumber\\
\!\!\!&=&\!\!\!\sum_{i=1}^{N} \left<r_i^2\mu_i^2\right>+\sum_{\substack{i,j\\i\neq j}}^{N}\left<r_i r_j\mu_i\mu_j\right>
\end{eqnarray}
Here, $r_i$ is the magnitude of the $i$th displacement $\mathbf{r}_i$ (i.e., $r_i=\left|\mathbf{r}_i\right|$), and $\mu_i=\cos\theta_i$ is the cosine of the polar angle of $\mathbf{r}_i$, measured from the positive $z$-axis direction, as illustrated in Figure \ref{fig01}. The angles $\theta_i$ and $\theta_j$ are independent, and thus the second term will vanish if the scattering is front-back symmetric. The first term becomes $N_{\rm scatt}\left<r^2\mu^2\right>$ if photons undergo scattering $N_{\rm scatt}$ times before escaping. Here, $\left<r^2\mu^2\right>$ represents the mean squared $z$ coordinate of the displacement vector between scattering events. The radiation field would be isotropic in the optically thick limit, even when the scattering is not isotropic, if it is front-back symmetric. In that case, the polar angle $\mu$ will be uniformly distributed, and $\left<r^2\mu^2\right>$ can be estimated as follows:
\begin{eqnarray}
\left<z^2\right>=\left<r^2\mu^2\right>\!\!\!&=&\!\!\!\frac{1}{2}\int_{-1}^{1}\int_{0}^{\infty}\left(\tau \ell\right)^2 \mu^2 e^{-\tau}d\tau d\mu \nonumber\\
\!\!\!&=&\!\!\!\frac{2}{3}\ell^2
\end{eqnarray}
As a result, the mean number of scatterings can be obtained from the condition $N_{\rm scatt}\left<r^2\mu^2\right>=L^2$, as follows:
\begin{equation}
	N_{\rm scatt}\approx \frac{3}{2}\tau_0^2~~~~(\tau_0\gg 1,~{\rm slab})
\end{equation}

As for the spherical geometry, it is tempting to add the equations for the optically thin and thick case. However, it was found that the equation for the optically thin slab is applicable only to much lower optical thickness ($\tau_0\lesssim 5\times 10^{-2}$) compared to that for a sphere. This arise because the optical thickness of a slab along a direction parallel to the $xy$ plane is, in fact, infinite. The formula was also found to significantly deviate from the simulation results at intermediate optical depths. Therefore, instead of simply combining the formulas for optically thin and thick limits, we made slight modifications to Equation (\ref{eq-for-thin-slab}) for the optically thin case and obtained an approximate formula. The final approximate formula is given as follows:
\begin{equation}
N_{\rm scatt}^{\rm slab} \approx  \frac{\tau_0\left(1-\gamma-\ln\tau_0+5.6\tau_0\right)}{1+\tau_0^2} + \frac{3}{2}\tau_0^2.\label{eq-slab}	
\end{equation}
In the next section, we demonstrate that this adjusted formula reproduces the simulation results fairly well.

\begin{figure*}[t]
\centering
\includegraphics[angle=0.0,width=82mm]{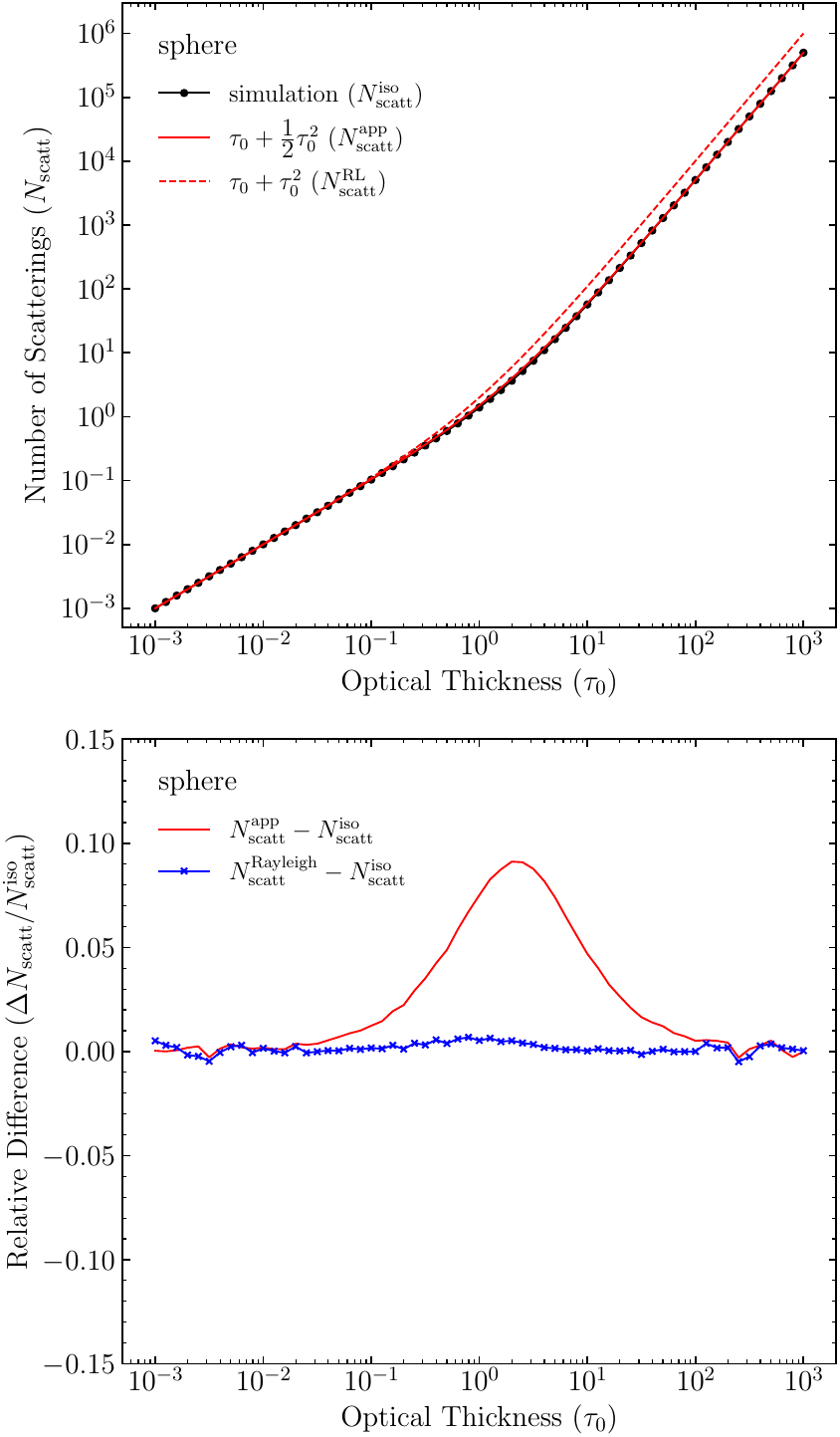}\hspace{4mm}
\includegraphics[angle=0.0,width=82mm]{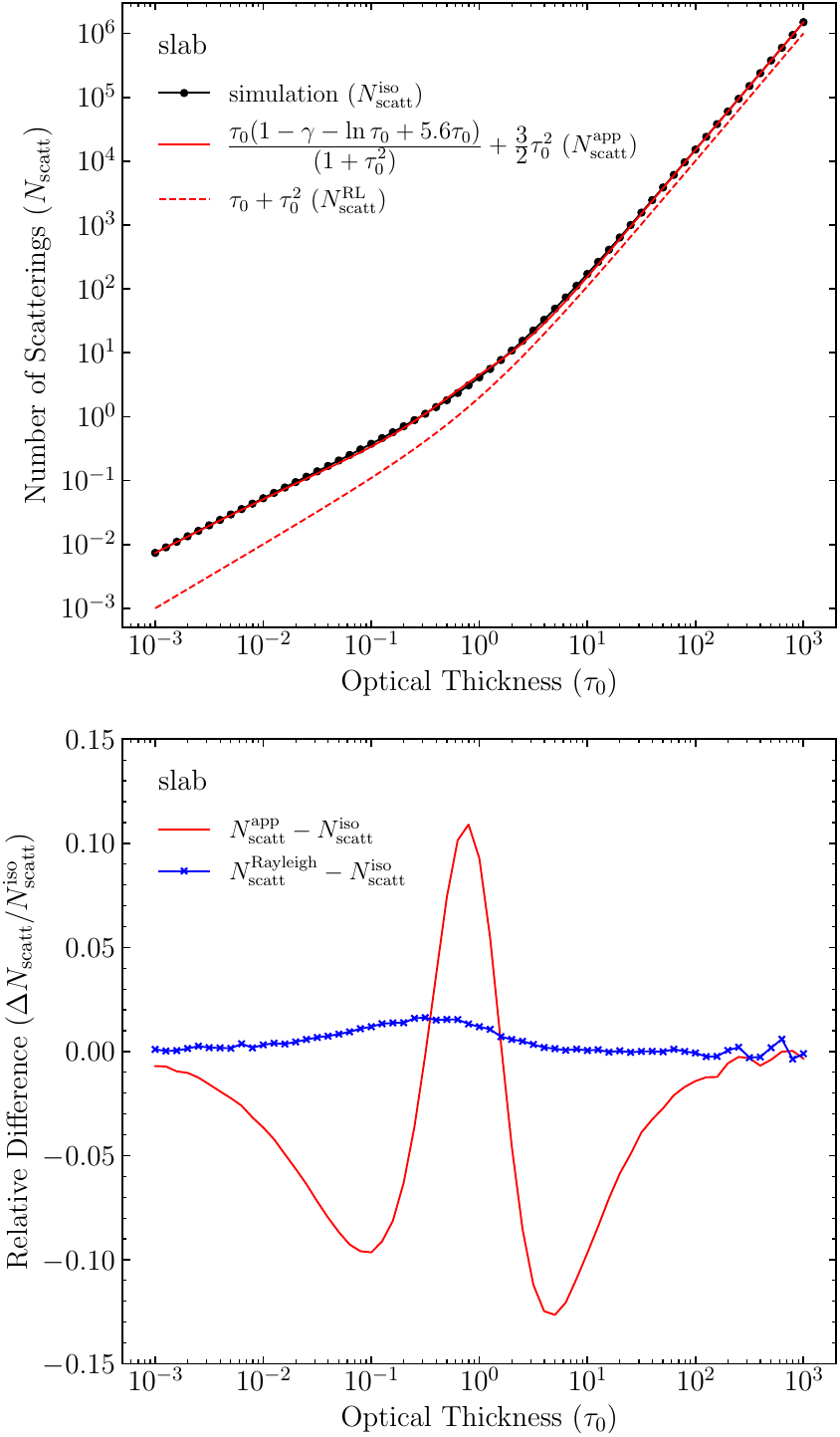}
\caption{Monte Carlo RT simulation results for the number of scatterings as a function of the optical thickness ($\tau_0$) in spherical and slab geometries. The left panels show the results for spherical geometry, and the right panels display those for slab geometry. The upper panels compare the number of scatterings with the analytical formulas, while the lower panels show the relative differences of the approximate formulas from those obtained in simulations. The lower panels also display the difference between the results for Thomson scattering and isotropic scattering. In the upper panels, the black lines with circles represent simulation results. Red solid and dashed lines denote the approximate formulas derived in this paper and in \citet{Rybicki1986_book}, respectively. In the figures, $N_{\rm scatt}^{\rm iso}$ and $N_{\rm scatt}^{\rm Rayleigh}$ denote the results obtained from Monte Carlo simulations, assuming isotropic and Thomson scattering phase functions, respectively. $N_{\rm scatt}^{\rm app}$ indicates the number of scatterings calculated using the analytic formulas derived in this work. $N_{\rm scatt}^{\rm RL}$ denotes the formula given by \citet{Rybicki1986_book}.\label{fig02}}
\end{figure*}
\section{Monte Carlo Simulation\label{sec:03}}
In this section, we describe our Monte Carlo RT simulation method and compare the simulation results with the analytical approximation equations for the number of scatterings.

\subsection{Simulation Methods}
The Monte Carlo simulation algorithms adopted in this study are similar to those described in \citet{Seon2020ApJS} and \citet{Choe2023}.
Photons are traced from the center, where they are generated, until they escape from the scattering region. The emitting source is assumed to be monochromatic and isotropic.
The scattering cross section is assumed to be independent of the wavelength, akin to the Thomson scattering process. The optical depth $\tau$ traveled by a photon between scattering events is directly proportional to the physical path length $\left|\mathbf{r}_i\right|$.

Simulations were performed using two different approaches: with and without adopting the first forced scattering algorithm. In the first approach, the optical depth traveled by a photon before experiencing the next scattering event is randomly chosen as follows:
\begin{equation}
\tau=-\ln\xi,
\end{equation}
where $\xi$ is a uniformly distributed random number between 0 and 1. In this approach, the photon weight is always $w=1$.
In the second approach, to minimize the inefficiency inherent in simple Monte Carlo simulations in the optically thin limit, the forced scattering algorithm is adopted for the first scattering event for every photon \citep[e.g.,][]{Cashwell1959_book,Witt1977,Baes2011,Seon2023}. The optical depth traveled between scattering events is randomly chosen as follows:
\begin{equation}
	\tau=-\ln\left(1-\xi w\right),
\end{equation}
where $\xi$ is a uniform random number, and $w$ is the photon weight. The photon weight is $w=1-e^{-\tau_{\rm max}}$ for the first scattering event, and $w=1$ for subsequent scatterings. Here, $\tau_{\rm max}=\tau_0$ for a sphere, while for a slab, $\tau_{\rm max}=\tau_0/\left|\mu\right|$ when the photon is emitted along a direction corresponding to $\mu$. The results obtained using both approaches were found to agree very well. In this paper, the results obtained using the first approach are presented.

The next scattering location from the current position $\mathbf{r}$ is determined by $\mathbf{r}'=\mathbf{r}+\tau\ell\mathbf{k}$, where $\mathbf{k}$ is the current propagation direction vector of the photon before the scattering. Because the scattering medium is assumed to have a constant density in this study, distances are measured in units of the mean free path (i.e., $\ell=1$).
After determining the next scattering location, the radial distance of the photon is compared with the radius in the case of spherical geometry or the $z$ coordinate is compared with the height of the medium in the case of slab geometry. The comparison determines the fate of the photon -- whether it escapes the system or continues scattering.

If the photon does not escape, the next scattering propagation vector $\mathbf{k}'$ is chosen by sampling the scattering polar angle $\vartheta$ and azimuthal angle $\varphi$ relative to $\mathbf{k}$. The scattering angles are defined in a coordinate system formed by three unit vectors: $\mathbf{e}_z=\mathbf{k}$, $\mathbf{e}_y=\hat{\mathbf{z}}\times\mathbf{k}/\left|\hat{\mathbf{z}}\times\mathbf{k}\right|$, and $\mathbf{e}_x=\mathbf{e}_y\times\mathbf{e}_z$, provided $\mathbf{k}$ is not parallel to $\hat{\mathbf{z}}$ (i.e., $\hat{\mathbf{z}}\times\mathbf{k}\neq 0$). If $\mathbf{k}$ is parallel to $\hat{\mathbf{z}}$, the three basis vectors are chosen to be $\mathbf{e}_x=\hat{\mathbf{x}}$, $\mathbf{e}_y=\hat{\mathbf{y}}$, and $\mathbf{e}_z=\hat{\mathbf{z}}$. Once the scattering angles ($\vartheta$, $\varphi$) were sampled according to an appropriate PDF, the next propagation vector is obtained as follows:
\begin{equation}
	\mathbf{k}' = \sqrt{1-{\text\textmu^2}}\left(\cos\varphi\,\mathbf{e}_x +\sin\varphi\,\mathbf{e}_y\right) + {\text\textmu}\,\mathbf{e}_z,
\end{equation}
where ${\text\textmu}=\cos\vartheta$. The formulas given in \citet{Pozdnyakov1983} and \citet{Seon2009} are equivalent to the above equation, except for the definition of $\varphi$, which is modified as $\varphi\rightarrow\varphi-\pi/2$. In the case of isotropic scattering, ${\text\textmu}=2\xi_1-1$ and $\varphi=2\pi\xi_2$ for two independent, uniformly distributed random numbers $\xi_1$ and $\xi_2$.
In Thomson scattering, the scattering angle $\vartheta$ is sampled according to the distribution function proportional to $1+\cos^2\vartheta$. For Thomson scattering, $\text\textmu$ is chosen as follow for a uniform random number $\xi$:
\begin{eqnarray}
	Q \!\!\!&=&\!\!\! \left[2(2\xi-1)+\sqrt{4(2\xi-1)^2+1}\right]^{1/3}\nonumber\\
	{\text\textmu} \!\!\!&=&\!\!\! Q - 1/Q
\end{eqnarray}
This equation is derived in \citet{Seon2006} and further discussed in \citet{Seon2020ApJS}.

The number of scatterings for each photon is calculated by adding the photon weight $w$ every time it undergoes scattering. The process continues until the photon escapes. Finally, the mean number of scatterings is calculated by averaging the results obtained for a large number of photons. The number of photons used in this study ranges from $N_{\rm photons}=10^4$ to $10^8$, depending on the medium's optical depth $\tau_0$. For optically thin cases ($\tau_0\leq 10$), $10^7$ or $10^8$ photons were used, while a relatively small number of photons were adopted for optically thick cases.

\subsection{Comparison with the analytic approximations}
Figure \ref{fig02} presents the results of the Monte Carlo RT simulations conducted in a sphere (left panel) and in a slab (right panel). In the upper panels, the results obtained by assuming the isotropic scattering are compared with the analytical approximations, Equations (\ref{eq-sphere}) and (\ref{eq-slab}). Additionally, the formula proposed by \citet{Rybicki1986_book} is also compared. In the figure, the  black lines with circles represent the simulation results, while the red lines indicate the analytic approximations. The red solid line represents our formula, while the red dashed line represents the formula from \citet{Rybicki1986_book}. As can be seen in the figure, the simulation results are well reproduced using our formulas in both a sphere and a slab. On the contrary, in optically thick cases, the formula of \citet{Rybicki1986_book} overpredicts the simulation for a sphere by a factor of 2, while it underpredicts those for a slab by a factor of 1.5. In optically thin cases, it significantly underpredicts the results for a slab. It is also evident that, for a given optical thickness $\tau_0$, the number of scatterings in a slab is always higher than that in a sphere, as the optical thickness in a slab depends on the photon propagation direction and becomes very high in the $xy$ plane direction.

The red lines in the lower panels present the relative discrepancies between the approximate formulas and the simulations where the isotropic scattering is adopted. The figures illustrate that the approximate formulas reasonably account for the simulation results within approximately 9\% for a sphere and 13\% for a slab. It is evident that the accuracy of the approximate formula for a sphere is better than that for a slab. This arises from the fact the optical thickness toward the direction near the $xy$ plane in a slab can be very large even in the case of $\tau_0<1$.

The lower panels in Figure \ref{fig02} also display the relative differences between the simulation results for isotropic scattering and Thomson scattering in the blue lines with crosses. Although the differences are minor, one might recognize that the Thompson scattering results yield slightly, but systematically, larger numbers of scatterings, particularly in the case of a slab geometry with optical thicknesses of $0.1\lesssim\tau_0\lesssim 1$. This minor difference is attributed to the fact that, in relatively low optical thicknesses, the radiation field in an asymmetric slab is less isotropic compared to the spherical symmetric case. However, at high optical thicknesses ($\tau_0\gtrsim 1$), the radiation field rapidly become isotropic even in a slab geometry.

\section{Summary and Discussion\label{sec:04}}
In this paper, we derived approximate formulas (Equations \ref{eq-sphere} and \ref{eq-slab}) for the number of scatterings in random walk processes, where the cross section is independent of the photon frequency, both in a sphere and a slab. The formulas demonstrate a good match with the Monte Carlo simulation results, with discrepancies reaching a maximum of only approximately 9\% for a sphere and 13\% for a slab. The formulas apply to both isotropic and Thomson scattering.

It is also found that, for a given optical thickness, the number of scatterings in a slab is always higher than that in a sphere. This is because the optical thickness of photons propagating approximately parallel to the $xy$ plane in the slab can be very high.

It is noteworthy that in the optically thick limit, the number of scatterings in Ly$\alpha$ RT is proportional to $\tau_0$: $N_{\rm scatt}=0.9579\tau_0$ for a sphere and $N_{\rm scatt}=1.612\tau_0$ for a slab \citep{Seon2020ApJS}. This linear dependence is significantly different from that expected in random walks, where the number of scatterings is proportional to $\tau_0^2$. This difference arises from the diffusion in frequency that occurs in Ly$\alpha$ RT, a phenomenon not present in random walks. Once the frequency of a Ly$\alpha$ photon diffuses into the wings far from the line center, the cross section decreases dramatically, leading to escape through a single longest `excursion,' as described by \citet{Adams1972}. Consequently, the number of scatterings decreases markedly compared to the case of random walks.

While it is beyond the scope of this paper to investigate line broadening in detail based on the number of scatterings, it is reasonable to assume that, after a sufficiently large number of scatterings, the core of an emission line profile will be well described by a Gaussian profile, with its width approximately proportional to the square root of the number of scatterings (equivalently, proportional to the optical depth), owing to the independence of the scattering cross-section on the photon frequency. In Figure 6 of \citet{Choe2023}, the full width at half maximum (FWHM) of the Thomson-scattered line for $\tau=10$ is approximately twice as wide as that for $\tau=5$, consistent with our expectation. Therefore, in the case of spherical geometries, using the formula from \citet{Rybicki1986_book} would predict a broader line width by a factor of $\sqrt{2}$ for a given optical depth (column density). Conversely, for slab geometries, the line width is predicted to be narrower by a factor of $\sqrt{1.5}$ at a fixed optical depth. In practical terms, when analyzing observational data, their formula will suggest a lower column density by a factor of $\sqrt{2}$ in spherical geometry and a higher column density by a factor of $\sqrt{1.5}$ in slab geometry.

In the present study, we considered uniform density media. In a plane-parallel medium, where the density depends only on the height ($\left|z\right|$) from the midplane, the number of scatterings is expected to be independent of the detailed density dependence on $\left|z\right|$. This independence arises because the integral of a density $n(z)$ over a length $\Delta s = \Delta z/\sin\left|b\right|$ along a direction angle $b$, measured from the $xy$ plane, is simply given by $\int n(z)dz/\sin\left|b\right|$. This integral is independent of the detailed functional form of $n(z)$. Indeed, in their Ly$\alpha$ radiative transfer study in media with power-law density profiles, \citet{Lao_Smith2020} found that emergent Ly$\alpha$ spectral profiles from any density profile in slab geometry are equivalent to those from a uniform medium with the same vertically-measured optical depth.
However, in sphere geometry, the integral of a density profile $n(r)$ -- a function of the radius $r$ -- over a distance along a direction strongly depends on the detailed structure of the density profile along the path. In the study by \citet{Lao_Smith2020}, it was found that the emergent Ly$\alpha$ spectrum from a sphere with a power-law density profile ($n\propto r^{-\alpha}$ for a positive $\alpha$) becomes narrower, and its peak shifts towards the line center as the density drops more steeply (larger $\alpha$). This result indicates that the number of scatterings of Ly$\alpha$ photons is smaller for a steeper density profile, characterized by a higher density at the center. Similar trends are expected in the case, where the cross-section is independent of the frequency, as addressed in the present study. However, no detailed interpretation of this behavior was provided in their study. This trend is attributable to the fact that the solid angle subtended by a high-density central region from a location is always smaller than that corresponding to the same column density expected in the case of uniform density. Consequently, the probability of scattering toward low densities is higher than in a uniform sphere, resulting in a smaller number of scatterings.


\acknowledgments
The authors are grateful to the referee, who provided constructive comments. This work was partially supported by a National Research Foundation of Korea (NRF) grant funded by the Korea government (MSIT) (No. 2020R1A2C1005788) and by the Korea Astronomy and Space Science Institute grant funded by the Korea government (MSIT; No. 2023183000 and 2023186903).

\end{document}